\documentclass[letterpaper, 10 pt, conference]{ieeeconf}  

\pdfoutput=1

\IEEEoverridecommandlockouts                              

\usepackage{amsmath, amsthm, amsfonts, amssymb,color}
\usepackage[ruled,vlined,noend]{algorithm2e}
\usepackage{blindtext}
\usepackage{graphicx}
\usepackage[utf8]{inputenc}
\usepackage{mathtools}
\usepackage{tikz, cite}
\usepackage[font=footnotesize,labelformat=simple]{subcaption}
\usepackage{tabu}

\newtheorem{example}{Example}
\usepackage{epsfig} 
\usepackage{psfrag}
\usepackage{mathrsfs}

\usepackage{multirow}
\usepackage{array,booktabs,arydshln,xcolor}
\usepackage{soul}
\usepackage{mathtools}
\usepackage{nccmath}
\usepackage{caption}
\usepackage{nccmath}
\usepackage{comment}
\usepackage{pdfpages}
\usepackage{comment}
\renewcommand{\vec}[1]{\boldsymbol{#1}}
\usepackage{float}
\usepackage{graphicx}
\usepackage{epstopdf}
\usepackage{color}
\usepackage{verbatim}
\usepackage{tikz,times}
\usepackage{xcolor}
\usepackage{lipsum}
\usepackage{makecell}
\let\labelindent\relax

\usepackage{enumitem}
\usepackage{multirow}

\usepackage{stfloats}

\title{\LARGE \bf
Reachability-Based Contingency Planning  against Multi-Modal Predictions with Branch MPC
 }

\author{Mohamed-Khalil Bouzidi$^{1,3}$, Bojan Derajic $^{2,3}$,  Daniel Goehring$^{1}$, Joerg Reichardt$^{3}$
\thanks{$^{1}$  Freie Universität Berlin, Germany }
\thanks{{\tt\small \{firstname.lastname@fu-berlin.de\}}}
\thanks{$^{2}$ Technische Universität Berlin, Germany }
\thanks{{\tt\small \{firstname.lastname@tu-berlin.de\}}}
\thanks{$^{3}$ Continental AG}
\thanks{{\tt\small \{firstname.lastname@continental.com\}}}
\thanks{This work is funded by the German Federal Ministry for}
\thanks{Economic Affairs and Climate Action within the project "nxtAIM".}
}

\begin{document}

\maketitle
\thispagestyle{empty}
\pagestyle{empty}

\begin{abstract}
This paper presents a novel contingency planning framework that integrates learning-based multi-modal predictions of traffic participants into Branch Model Predictive Control (MPC). 
Leveraging reachability analysis, we address the computational challenges associated with Branch MPC by organizing the multitude of predictions into driving corridors. Analyzing the overlap between these corridors, their number can be reduced through pruning and clustering while ensuring safety since all prediction modes are preserved.
These processed corridors directly correspond to the distinct branches of the scenario tree and provide an efficient constraint representation for the Branch MPC. We further utilize the reachability for determining maximum feasible decision postponing times, ensuring that branching decisions remain executable. Qualitative and quantitative evaluations demonstrate significantly reduced computational complexity and enhanced safety and comfort.
\end{abstract}
\section{Introduction} \label{sec:Intro}
Safety and comfort in autonomous vehicles (AVs) increase with longer planning horizons. With longer planning horizons, the necessary anticipation of other traffic participants' (TPs') motion, however, becomes increasingly difficult and uncertain. Often, several quite different, but similarly plausible options may exist for the future trajectory of the TPs surrounding an AV. In recent years, learning-based predictors have started to address this challenge of multi-modality in the distribution of future TP trajectories \cite{nayakanti_wayformer_2022, shi_motion_2022, yue, Schlauch_eai} outperforming  traditional methods in accuracy. 

Making effective use of multi-modal predictions for planning tasks, i.e., acting in a way that is compatible with a multitude of plausible evolutions of a traffic scene, however, remains an open challenge. Traditional approaches plan only for the most likely prediction, which proves inadequate in safety-critical applications. Robust\cite{singh, batkovic} and stochastic \cite{benciolini_multistage_2021, brudigam_combining_2018, kensbock_scenario-based_2023} approaches, which generate a single plan by considering all or most scenarios simultaneously over the entire planning horizon, often become overly conservative. They fail to account for the reduction in uncertainty as new information becomes available over time. Branch Model Predictive Control (MPC)\cite{ chen_interactive_2021, zhangbmpc, fors_resilient_2022,  ulfsjoo_integrating_2022, bouzidi2024motionplanninguncertaintyintegrating} and other Contingency Planner \cite{khaled, peters, Cui_2021_ICCV} have emerged as a promising alternatives that allow different plans to coexist as long as they are identical during a certain number of initial timesteps. They effectively postpone decisions until uncertainty is reduced. While principled, these approaches face significant scaling challenges with increasing numbers of traffic participants and scenarios in their scenario tree.

\begin{figure}[t]
\centering
    \includegraphics[trim = 10mm 27mm 34mm 4mm, width=0.49\textwidth]{"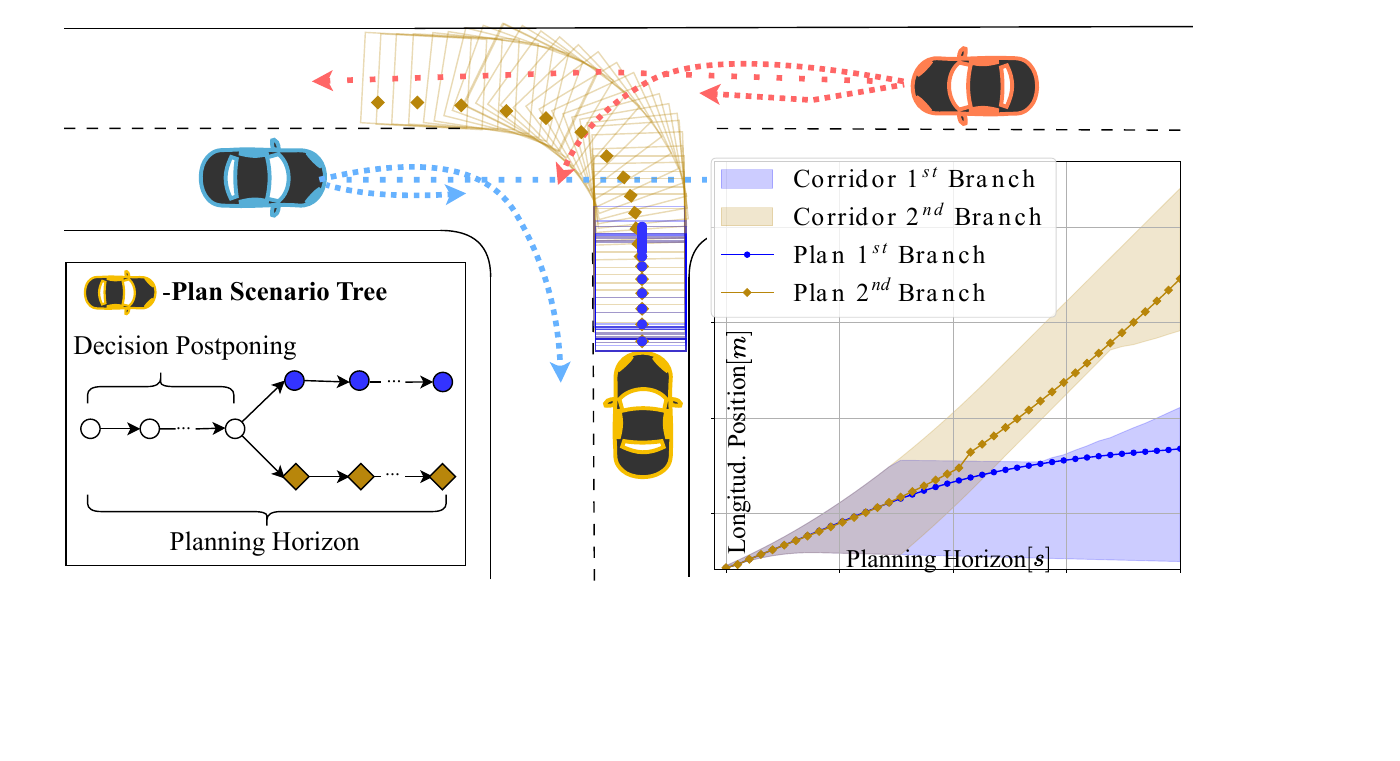"}
    \caption{Illustration of planning in traffic scenes with uncertain multi-modal predictions: Mode-dependent driving corridors are extracted to construct the Branch MPC scenario tree.}
    \label{fig:P1}
\vspace{-1.5em}    
\end{figure}

A natural strategy to address this scaling problem is to reduce the number of predicted modes. Na\"ively considering only a fixed number of modes with highest probability can be misleading as as even lower probability predictions can represent meaningful behaviors, making their exclusion a potential safety risk. On the other hand, only considering the “worst-case” scenarios leads to overcautious behavior.

Also, more sophisticated methods  for scenario selection \cite{bouzidi2024motionplanninguncertaintyintegrating, ulfsjoo_integrating_2022, fors_resilient_2022} do not guarantee that the pruned scenarios are satisfied in Branch MPC,  potentially compromising safety capabilities. The approaches \cite{bouzidi2024motionplanninguncertaintyintegrating,fors_resilient_2022} prunes scenarios based on the previously computed optimal trajectory, assuming minimal changes between timesteps. This assumption may break down in uncertain environments, potentially omitting safety-critical predictions. The work in \cite{ulfsjoo_integrating_2022} carries the same risk as it employs a similarity measure to compare and prune scenarios with minimal differences without accounting for the safety criticality of eliminated scenarios. Moreover, they fail to address the scaling problem as TP numbers increase.

In this work, we propose a novel approach that addresses these challenges by extracting multiple driving corridors for the AV for each predicted mode based on reachability analysis. With Reachability analysis, which is already successfully used for planning \cite{reach, reach1, reach2},  we can calculate the set of all possible future states the AV can reach, considering its initial state, dynamics, constraints, and obstacles. Our method merges these reachable sets/corridors with significant overlap while maintaining separate scenarios for those with minimal overlap, efficiently managing the scenario tree while preserving safety (see Fig \ref{fig:P1}). 
We further leverage reachability analysis to address the branching time calculation in Branch MPC. Recent works\cite{ bouzidi2024motionplanninguncertaintyintegrating,peters, khaled} have concluded that an appropriately selected branching time is crucial, and while several heuristics have been proposed, the calculation of an optimal branching time still is an unresolved question. Too short branching times can lead to premature commitment to scenarios under high uncertainty, while excessive branching times may result in missed opportunities for promising maneuvers. By leveraging the calculated reachable sets, we determine the maximum possible branching time during which the AV can stay within all driving corridors in the scenario until the end of the planning horizon. This allows us to either constrain the estimated decision postponing time or replacing overly restrictive driving corridors.

In summary, this paper makes the following contributions:
\begin{itemize}[leftmargin=*]
    \item  A contingency planning framework that integrates learning-based multi-modal predictors in Branch MPC
    \item A safe and efficient reachability-based scenario tree construction and constraint handling approach for Branch MPC
    \item A novel method for incorporating maximum feasible decision postponing through reachability analysis
\end{itemize}


\section{Preliminaries} \label{sec:pre}

We consider an arbitrary traffic scene comprising $O$ Traffic participants $o \in \{0,..,O-1\}$. 
The problem setup incorporates a map $\mathbf{M}_{in}$ of the environment and a reference path $\mathcal{P}_{ref}: [0,\theta_{max}] \to \mathbb{R}^4 \times [0, 2\pi]$ generated by an upstream route planner. This reference path maps the arclength $\theta$ to the tuple $\theta \mapsto (x_{ref}(\theta), y_{ref}(\theta), d_{lb}(\theta), d_{rb}(\theta), \psi_{ref}(\theta))$, comprising positional coordinates $(x_{ref}, y_{ref})$, lateral distances to the road boundaries $d_{lb}, d_{rb}$, and a reference heading angle $\psi_{ref}$. The objective is to determine a collision-free AV trajectory along $\mathcal{P}_{ref}$.  This can be formulated as an optimal control problem (OCP):

\begin{equation} \label{eqn:J} \begin{aligned} \underset{\vec{z}_{k},\vec{u}_{k} \in [0,N]}{\min} \ & \sum_{k=0}^{N-1} J_k(\vec{z}_k, \vec{u}_k) + J_N(\vec{z}_N) \\ & s.t. \ \vec{z}_{k+1} = f(\vec{z}_{k},\vec{u}_{k}) \\ & \vec{z}_0 = \vec{z}(0) \\ & \vec{z}_{k} \in \mathcal{Z}, \ \vec{u}_{k} \in \mathcal{U} \end{aligned} \end{equation}  
where $J_k$ is the running cost, $J_N$ is the terminal cost and $N$ is the planning horizon.
The vehicle behavior in the OCP is approximated by discretizing the kinematic bicycle model:
\begin{equation}\label{eqn:f}  \dot{\vec{z}} = \begin{bmatrix} v \cos(\psi) \ v \sin(\psi) \ v \frac{\tan(\delta)}{l} \ a \ j \ \dot{\delta} \end{bmatrix} \end{equation}
The state vector $\vec{z}$ consists of the vehicle pose $(x, y, \psi)$, velocity $v$, acceleration $a$, steering angle $\delta$. The control input vector $\vec{u}$ includes the jerk $j$ and steering angle rate $\dot{\delta}$.
Further, the state constraints $\mathcal{Z}$ and the control input constraint $\mathcal{U}$ include actuator and dynamic limits \cite{polack}, road boundaries, and collision avoidance constraints. The collision constraints depend on the future states of the TPs, represented as $\vec{o}^o_k = [x^o_{k}, y^o_{k}, \psi^o_{k}, v^o_{k}]^\top$. As the intentions of other TPs are unobservable, these states must be predicted.

\subsection{Learning-based Multi-Modal Motion Predictor} \label{sec:predictor}
The inherent uncertainty in the intentions of the TPs requires a probabilistic prediction approach. Learning-based multi-modal prediction approaches have emerged as state-of-the-art solution for this challenge, trained on large-scale datasets such as Waymo Open \cite{ettinger_waymo_2021}. These approaches generate multi-modal predictions that capture different potential future behaviors by processing the logged sequence of the TP history states $\vec{o}_k$ and the road map $\mathbf{M}_{in}$.
For our {framework}\footnote{Our framework is compatible with other predictors that provides multi-modal predictions}, we employ the Motion Transformer \cite{shi_motion_2022} - a neural network with transformer-based encoder-decoder architecture. It outputs predictions in the form of a Gaussian Mixture Model (GMM) with 
\begin{equation}\label{eq:gmm} p^o_{k} = \sum_{m \in \mathcal{M}} \pi^o_m \mathcal{N}^o_{k,m}(\vec{\mu}^o_{k,m}, \boldsymbol{\Sigma}^o_{k,m}) \end{equation}
where each mixture component $m$ represents a distinct predicted trajectory mode with probability $\pi^o_m$, and predicted states $\vec{o}^o_{k,m} =\vec{\mu}^o_{k,m}= [x^o_{k,m}, y^o_{k,m}, \psi^o_{k,m}, v^o_{k,m}]^\top$ over the prediction horizon $N$.

\subsection{Nominal Model Predictive Contouring Control}\label{sec:nominal}
We use Model Predictive Contouring Control (MPCC) \cite{brito1} as the baseline and backbone of our framework. Conventional MPCC handles only one prediction per TP (typically the most likely one).

For the MPCC, the cost in eq.(\ref{eqn:J}), is composed of a term for minimizing lag error $\hat{e}^l_k$, contouring error $\hat{e}^c_k$, and control inputs and maximizing the virtual speed along the path denoted as $v^p_k$. This is expressed in the following cost function:
\begin{equation}
J_k = \left[\begin{array}{c}
\hat{e}^c_k \\
\hat{e}^l_k
\end{array}\right]^\top \mathbf{Q} \left[\begin{array}{c}
\hat{e}^c_k \\
\hat{e}^l_k
\end{array}\right] - q_v v_k^p + \vec{u}_k^\top \mathbf{R} \vec{u}_k 
\end{equation}
where $\mathbf{Q}$, $q_v$, and $\mathbf{R}$ are the respective weights. For this formulation, the system dynamics in eq. (\ref{eqn:f}) is extended by arclength $\theta_k$ as a state and control input $v_k^p$:
\begin{equation}\label{eq:theta}
\theta_{k+1} = \theta_k + v_k^p\Delta t
\end{equation}
where $\Delta t$ is the sampling time. Using this, we can approximate the contouring and lag errors:
\[
\left[\begin{array}{c}
\hat{e}^c_k \\
\hat{e}^l_k
\end{array}\right] = \left[\begin{array}{cc}
\sin(\psi_{\text{ref}}(\theta_k)) & -\cos(\psi_{\text{ref}}(\theta_k)) \\
-\cos(\psi_{\text{ref}}(\theta_k)) & -\sin(\psi_{\text{ref}}(\theta_k))
\end{array}\right] \Delta \vec{p}_{\text{ref}}
\]
where $\Delta \vec{p}_{\text{ref}} = [x - x_{\text{ref}}(\theta_k), y - y_{\text{ref}}(\theta_k)]^\top$.

Additionally, we extend the cost by applying a potential field around the TP and lane markers as described in \cite{bouzidi2024motionplanninguncertaintyintegrating}.

\begin{figure*}[t]
\vspace{0.6em} 
    \centering
    \includegraphics[trim = 0mm 30mm 0mm 0mm, width=1\textwidth]{"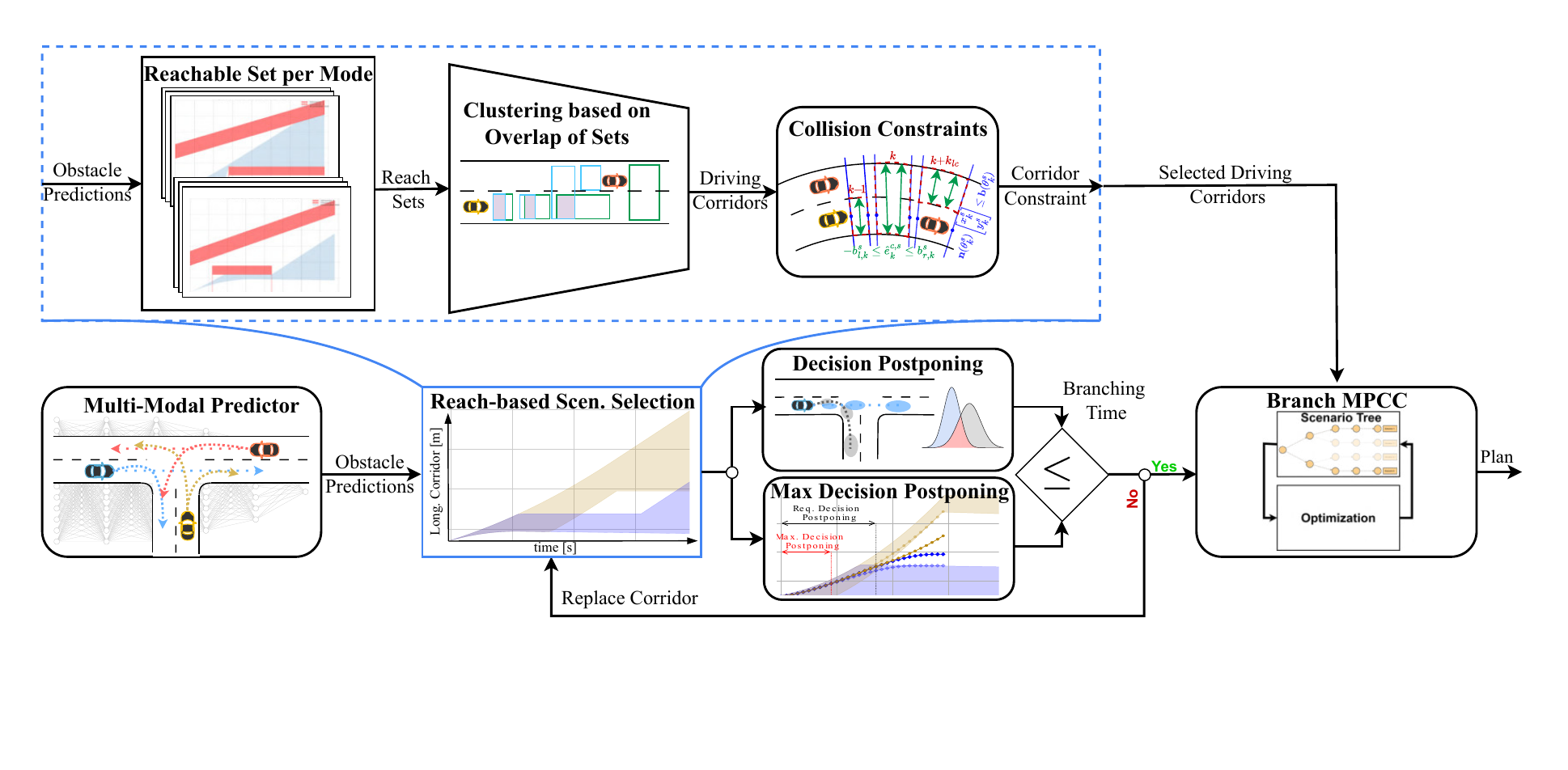"}
    \caption{Our Contingency Planning Framework  using a Multi-modal Predictor, Reachability-based Scenario Selection, Adaptive Decision Postponing and Branch MPCC}
    \label{fig:method}
    \vspace{-1.2em} 
\end{figure*}
\section{Branch Model Predictive Contouring Control} \label{sec:BMPCC}

Branch MPCC extends the conventional MPCC framework to handle multi-modal predictions through a scenario tree structure (s. fig \ref{fig:P1}).  Every branch in the scenario tree represents one distinct realization of the uncertainty.  In our case, this realization corresponds to the collision constraints with respect to the multi-modal TP predictions. The details on how these modes contribute to building the scenario tree will be introduced in sec. \ref{eq:reach}.
Based on that, the Branch MPCC generates a trajectory tree, i.e., contingency plans against each realization. We consider only a single branching point into $|\mathcal{S}|$ different scenarios. Each scenario $s\in\mathcal{S}$ consists of a sequence of $N$ nodes extending from the root to a leaf node.

Each of theses branches $s$ maintains its own state trajectory $\vec{z}_{k,s} \in \mathbb{R}^{n_z}$ and control inputs $\vec{u}_{k,s} \in \mathbb{R}^{n_u}$, where $n_z$ and $n_u$ denote the state and control dimensions respectively. As each node represents one state, control input pair the system dynamics evolve according to $\vec{z}_{k+1,s}= f(\vec{z}_{k,s},\vec{u}_{k,s})$.

Defining the set $I$ to contain all nodes $(k,s)$ within the scenario tree, enables the formulation of the optimization problem over the entire tree:
\begin{subequations}
\label{eq:NLP}
\begin{align}
 &\underset{\vec{z}_{k,s},\vec{u}_{k,s} \forall (k,s) \in I}{\min} \  \sum_{s \in\mathcal{S }}\pi_s \sum_{k=0}^{N-1} J_k(\vec{z}_{k,s}, \vec{u}_{k,s})  \\
 & s.t.  \   \vec{z}_{k+1,s}= f(\vec{z}_{k,s},\vec{u}_{k,s}) \ \forall (k,s) \in I  \\
 & \vec{z}_{0,s} = \vec{z}(0) \ \forall s \in \mathcal{S} \\
 & \vec{z}_{k,s} \in  \mathcal{Z}, \ \vec{u}_{k,s} \in \mathcal{U} \ \forall (k,s) \in I \\
 & \vec{u}_{k,i} = \vec{u}_{k,j} \ \forall (i,j)\in \mathcal{S} \  \forall \ k \in [0, k_b]
\end{align}
\end{subequations}

employing the running cost and constraint from the nominal MPCC as defined in sec. \ref{sec:nominal}.
The branch weight $\pi_s$ often corresponds to the occurrence probability of the respective scenario.

A key feature of Branch MPC is the incorporation of non-anticipatory constraints, expressed in equation (\ref{eq:NLP}e). These constraints enforce identical control inputs across all scenarios during the initial timesteps $[0, k_b]$, where $k_b$ represents the branching time or decision postponing time. This leads to passive information gathering, acknowledging that the AV does not know initially which scenario will unfold and that  new information will emerge over time. Consequently, the AV follows a consensus plan satisfying all scenario constraints until $k_b$, postponing commitment to any scenario before branching and adapting its strategy according to the realized scenario. We detail the choice of $k_b$ in sec. \ref{sec:adp}.

\section{Reachability-based Scenario Selection}\label{sec:scen}

This section outlines our methodology for constructing the scenario tree of the Branch MPC. We begin by employing reachability analysis to compute driving corridors corresponding to each predicted mode $m\in\mathcal{M}$ for all TPs obtained from the multi-modal predictor. While each driving corridor for a predicted mode could theoretically define constraints for a separate branch, this approach would generate an excessive number of branches, resulting in high computational costs. To address this, we introduce a pruning and merging strategy to reduce the complexity of the extracted driving corridors. Finally, we detail the process of integrating the resulting constraints into the Branch MPC.

\subsection{Driving Corridor Extraction} \label{sec:dc}
\begin{figure}[t]
\centering
    \includegraphics[trim = 14mm 3mm 6mm 0mm, clip, width=0.49\textwidth]{"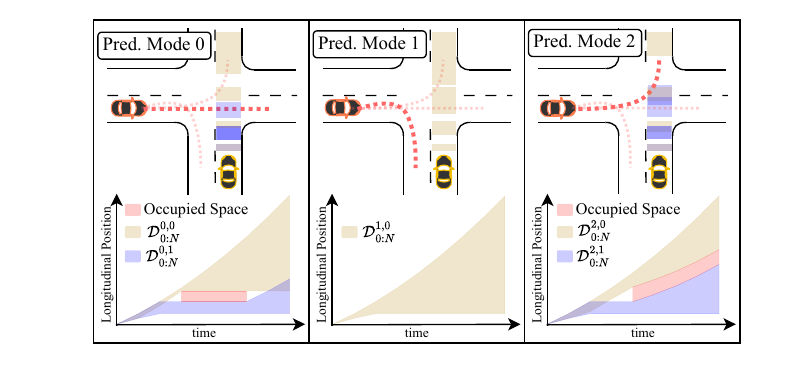"}  
    \caption{For each predicted mode of the TP, at least one safe driving corridor is identified. The upper plots show the respective corridors in a bird's-eye view, while the lower plots illustrate them in a path-time diagram. }
    \label{fig:P2} 
    \vspace{-1.3em} 
\end{figure}

Given the initial AV state $\vec{z}_0$ and the forbidden set of states $\mathcal{F}^m_k$ (e.g. the occupied space of the predicted states of the TP of a single-mode $m$) the reachable set $\mathcal{R}^m_k$ for each timestep  can be computed. $\mathcal{R}^m_k$ represents the set of all possible future states the vehicle can reach at timestep $k$, considering actuator limits and the occupied space. Exact reachability analysis is computationally intractable \cite{reach1}. Hence, we approximate the reachable set for each timestep using zonotopes, which allow efficient set operations such as the Minkowski sum $\oplus$ to be performed. To approximate the longitudinal behavior of the AV, we use a simple point mass model in Fr\'enet coordinates along the reference path with the states $[\theta, v^p_{k}]^T $ using eq. (\ref{eq:theta}) in the MPCC where we approximate $ v^p_{k+1}\approx v_{k+1}= v_k + a_k\Delta t$. Acceleration limits $[a_{min}, a_{max} ]$ and velocity limits $[0, v_{max}]$ are considered. The lateral behavior is modeled as discrete lane-change events, assuming a minimum lane-change time $k_{lc} = \sqrt{4 d_l / a_{max}}/\Delta t $ as in \cite{reach} where $d_l$ is the distance to the neighbouring lane. Hence, for a predicted mode $m$ and a given lane, the reachable set at timestep $k$ can be approximated by:
\begin{equation}\label{eq:reach}
    \mathcal{R}^{m,i}_{k+1} = 
    \left(
        \underbrace{\begin{bmatrix}
        1 &\Delta t \\
        0 & 1
        \end{bmatrix}}_{\vec{A}}
        \mathcal{R}^{m,i}_{k} \oplus
        \underbrace{\begin{bmatrix}
        0.5\Delta t^2 \\
       \Delta t
        \end{bmatrix}}_{\vec{B}}
        [a_{min}, a_{max}]
    \right)
    \setminus \mathcal{F}^m_{k+1}
\end{equation}
where $\mathcal{F}^m_k$ represents forbidden states such as obstacle-occupied space, velocity-limit violations, or lane endings. 
The superscript $i$ is introduced because the drivable area is typically non-convex, (i.e., it may have “holes” due to obstacles). Consequently, the reachable set $\mathcal{R}^{m}_{k}$ is decomposed into $d$ subsets $\mathcal{R}^{m}_{k} = \bigcup_{i=0}^{d-1} {R}^{m,i}_{k} $ , i.e.,  each representing a different possible maneuver (cf. fig \ref{fig:P2}).

Approximating reachable sets purely forward in time may still include states that are not truly reachable. One can refine these sets via backward propagation:
\begin{equation}\label{eq:brs}
    \mathcal{R}^{m,i}_{k-1}  \leftarrow \mathcal{R}^{m,i}_{k-1} \cap \vec{A}^{-1} \big( \mathcal{R}^{m,i}_{k} \oplus -\vec{B} [ a_{min}, a_{max} ] \big)
\end{equation}
However, to improve computational efficiency, we skip this backward step here and only apply it to the final selected corridors at the end of our approach.

Hence, we calculate each driving corridor $\mathcal{D}^{m,i}_{0:N}= \{ \mathcal{D}^{m,i}_k \mid k \in [0, N] \}$, by projecting the reachable set calculated in eq. (\ref{eq:reach}) onto the position domain $\mathcal{D}^{m,i}_k= \text{proj}_\theta(\mathcal{R}^{m,i}_{k}) \times [-b^{m,i}_{l,k}, b^{m,i}_{r,k}]$ , where $b^{m,i}_{l,k}$ and $b^{m,i}_{r,k}$ denote the distance from the reference to left and right boundary of the corridor. 

In single-lane scenarios, $b^{m,i}_{l,k}$ and $b^{m,i}_{r,k}$ can be derived directly from the lane width. For multi-lane scenarios, at each timestep, we check whether a lane change 
(requiring time \( t_{lc} \)) is feasible based on the TP predictions. Furthermore, we assume that in the given planning horizon, only one lane change is possible which is reasonable if the planning horizon $\leq 5s$.  Therefore, determining all possible driving corridors consists of computing the corridor for the current lane and for the adjacent lanes (left and right), yielding multiple subsets of the reachable set.
This procedure is repeated for each predicted $m \in \mathcal{M}$, as illustrated in fig. \ref{fig:P2} returning all possible driving corridors for each obstacle prediction of the multi-modal predictor.As computations across modes are independent, we parallelize their execution.

If the number of driving corridors per mode exceeds 1 (e.g.  mode 1 and mode 3 in fig. \ref{fig:P2}), it is customary to select the corridor with the greatest cumulative area, denoted as $\mathcal{D}^{m,A}_{0:N}$, since it generally leads to less restrictive position constraints for planning \cite{reach1}. The corridors which are not selected are kept as backups, whose usage is described in Sec. \ref{sec:adp}.

\subsection{Clustering and Pruning} \label{sec:iou}

After extracting one driving corridor for each predicted mode $\mathcal{D}^{m,A}_{0:N} \mid \forall m\in \mathcal{M}$ , we compare these driving corridors to each other to identify a small number of subsets that collectively satisfy all modes. This is achieved by evaluating the overlap between pairs of driving corridors and determining whether they should be merged based on a predefined similarity threshold.

For two driving corridor segments $\mathcal{D}^{m_1,A}_{k}$ and $\mathcal{D}^{m_2,A}_{k}$of two different modes $m_1, m_2$ at a given timestep $k$, the overlap metric \( \gamma_k \) is defined as Jaccard index:
\begin{equation}
    \gamma_k = \frac{\text{area}(\mathcal{D}^{m_1,A}_{k} \cap \mathcal{D}^{m_2,A}_{k})}{\text{area}(\mathcal{D}^{m_1,A}_{k} \cup \mathcal{D}^{m_2,A}_{k})}
\end{equation}

where \( \text{area}(\cdot) \) represents the area of the set. To evaluate the similarity of the two corridors over the entire planning horizon $N$, we compute the product of the overlap metrics across all timesteps $\Gamma = \prod_{k=0}^{N} \gamma_k$. A threshold $\Gamma_{\text{min}} > 0$ is introduced to control the clustering behavior. If $\Gamma \geq \Gamma_{\text{min}}$, the two driving corridors are merged into a single scenario. The resulting driving corridor is defined as the intersection of the original corridors:
\begin{equation}
\mathcal{D}^{\cap}_k = \mathcal{D}^{m_1,A}_{k} \cap \mathcal{D}^{m_2,A}_{k} \quad \forall k \in [0, N]  
\end{equation}
This ensures that the merged driving corridor satisfies all constraints from the original corridors, as it only includes states that are reachable in both.
If two corridors are merged we also sum up the respective probability of their modes and assign it to the new corridor.
Conversely, if $\Gamma < \Gamma_{\text{min}}$, the two corridors remain as separate clusters. This clustering approach ensures that driving corridors with small overlap, such as non-intersecting sets at any timestep, are treated as separate scenarios.
The parameter $ \Gamma_{\text{min}}$ balances conservatism and computational efficiency. A low threshold \( \Gamma_{\text{min}} \) leads to stronger merging, resulting in smaller and more restrictive corridors. This leads to more conservative vehicle behavior.  A high threshold \( \Gamma_{\text{min}} \) retains more distinct scenarios but increases the size of the scenario tree leading to higher computational complexity.

The final output is a reduced set of driving corridors
$
\{ \mathcal{D}^{s}_{0:N} \mid s \in \mathcal{S} \},
$
where each branch $s \in \mathcal{S}$ in the scenario tree has a corresponding driving corridor. These are subjected to backward propagation for refinement, as outlined in eq. (\ref{eq:brs}).The weight of each branch, \( \pi_s \), represents the assigned probability of its respective driving corridor. 
As long as $ \Gamma_{\text{min}} > 0 $, the resulting driving corridors collectively satisfy all predicted modes, with each mode $m \in \mathcal{M}$ accounted for in at least one $ \mathcal{D}^{s}_{0:N}$ for all $ k \in [0, N] $.

\subsection{Collision Constraints Formulation}\label{sec:constraint}
The driving corridors $\mathcal{D}^{s}_{0:N}$ which are added to the scenario tree  are represented in Frenet coordinates and need to be transformed into inequality constraints for the BMPCC. Here, we can leverage the property of the MPCC  that it naturally integrates progress along the reference path $\theta^s_k$ as an internal state and calculates the contouring error $e^{c,s}_k$ during optimization. This simplifies the formulation of the lateral and longitudinal constraints. The lateral boundaries of the driving corridor are enforced by directly constraining the contouring error at each timestep $k$ in the prediction horizon:
\begin{equation}
-b^{s}_{l,k} \leq e^{c,s}_k \leq b^{s}_{r,k}
\end{equation}
For the longitudinal constraints, we extract the minimum and maximum progress of the respective corridor $\theta^s_{k, \text{min}}, \theta^s_{k, \text{max}}$ at each timestep. Using these two values, the normal vector \( \vec{n}(\theta^s_k) \) to the reference path at \( (x_{\text{ref}}^s(\theta_k), y^s_{\text{ref}}(\theta_k)) \) is computed. These normal vectors define linear constraints on the \( x \)-\( y \)-coordinates for $\theta^s_{k, \text{min}}$ and $ \theta^s_{k, \text{max}}$:
\begin{equation}
\vec{n}(\theta^s_k) \cdot \begin{bmatrix} x^s_k \\ y^s_k \end{bmatrix} \leq \vec{b}(\theta^s_k)
\end{equation}
where $ \vec{b}$  ensures the AV remains within the corridor bounds.

\begin{figure}[t]
\centering
    \includegraphics[trim = 18mm 30mm 15mm 8mm, clip, width=0.45\textwidth]{"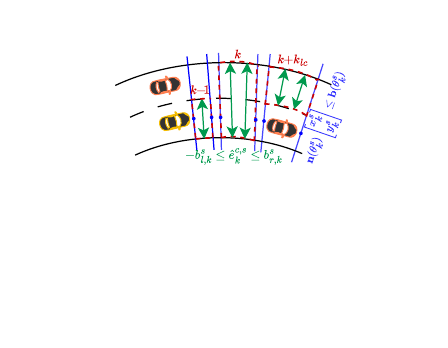"}  
    \caption{Illustration of the collision constraint formulation for respective driving corridor of one branch  $\mathcal{D}^s_{0:N}$}
    \label{fig:P3} 
    \vspace{-0.6em} 
\end{figure}

By combining these lateral and longitudinal constraints, we define a closed feasible area for the position in each timestep. On straight roads, these correspond to convex constraints. While convexity may be affected  by road curvature, the computational benefits remain significant. In typical formulations, collision constraints are inherently nonconvex. The AV is commonly approximated as multiple intersecting circles (e.g. 3 \cite{reach1}), while TPs are represented as ellipses. This introduces significant computational challenges for the solver. Moreover, such constraints scale poorly with the number of TPs, increasing the computational load. In the case of $|\mathcal{O}|$ TPs, the BMPCC in \cite{bouzidi2024motionplanninguncertaintyintegrating} leads to $3 \cdot |\mathcal{O}| \cdot N \cdot |\mathcal{S}|$ constraints for TPs and additionally $2 \cdot N \cdot |\mathcal{S}|$ constraints for road boundaries. In comparison our driving corridor constraints formulation is constant with respect to the number of TPs ($4 \cdot N \cdot |\mathcal{S}|$).

\begin{algorithm}[ht]
\caption{Contingency Planner}
\SetAlgoLined
\SetKwInOut{Input}{Input}
\SetKwInOut{Output}{Output}

\Input{
  Measured state values $ \vec{z}_{0}$, \\
  map information $\vec{M_{in}}$, reference path $\mathcal{P}_{ref}$, \\
  history states $ \vec{o}^o_{-\sigma:0}$ $\forall$ TPs $o \in \mathcal{O}$ 
}
\Output{Contingency Plans $\vec{z}_{0:N}$}

\textit{// Predict TP trajectories (Sec.~\ref{sec:predictor})} \\
$\vec{o}^o_{0:N,m}, \pi^o_m  \forall o \in \mathcal{O}, m \in \mathcal{M} \gets MTR(\vec{o}^o_{-\sigma:0}, \vec{M_{in}})$  \\

\textit{// Extract driving corridors (Sec.~\ref{sec:dc})} \\
\ForPar{$m \in \mathcal{M}$}{
        $\text{Append } \text{Reach}(\mathcal{P}_{\text{ref}}, \vec{o}^o_{0:N, m}, \text{Middle}) \text{ To } \mathcal{D}^{m}_{0:N}$ \\
        \If{left lane exists}{
            $\text{Append } \text{Reach}(\mathcal{P}_{\text{ref}}, \vec{o}^o_{0:N, m}, \text{Left}) \text{ To } \mathcal{D}^{m}_{0:N}$ \\
        }
        \If{right lane exists}{
                       $\text{Append } \text{Reach}(\mathcal{P}_{\text{ref}}, \vec{o}^o_{0:N, m}, \text{Right}) \text{ To } \mathcal{D}^{m}_{0:N}$ \\
        }
    
    $\mathcal{D}^{m,A}_{0:N} \gets \arg\max \text{Area}({D}^{m}_{0:N})$ \\
    $ \mathcal{D}^{m,B}_{0:N} \gets {D}^{m}_{0:N} \setminus \mathcal{D}^{m,A}_{0:N}$ \\

}


\While{Uncompared corridors exist}{
    \textit{// Cluster and prune corridors (Sec.~\ref{sec:iou})} \\
    \ForEach{pair of corridors $\mathcal{D}^{i,A}_{0:N}$, $\mathcal{D}^{j,A}_{0:N}$}{
        \If{$\Gamma(\mathcal{D}^{i,A}_{0:N}, \mathcal{D}^{j,A}_{0:N}) \geq \Gamma_{\text{min}}$}{
            $\mathcal{D}^{\cap}_k \gets \mathcal{D}^{i,A}_k \cap \mathcal{D}^{j,A}_k \ \forall k,  \pi_{\cap} \gets \pi_{i} + \pi_{j}$   \\
            Replace $\mathcal{D}^{i,A}_{0:N}$ and $\mathcal{D}^{j,A}_{0:N}$ with $\mathcal{D}^{\cap}_{0:N}$  \\
    }
    }

    \textit{// Max. Decision Postponing (Sec.~\ref{sec:adp})} \\
    $\mathcal{D}^{c,A}_{0:N}, k_b \gets \textbf{Alg.2}({D}^{c}_{0:N} \ \forall \text{ cluster c}, \mathcal{D}^{m,B}_{0:N} \ \forall m )$ \\

 } 
 $\{ \mathcal{D}^{s}_{0:N}, \pi_s | s \in \mathcal{S} \} \gets \text{Clustered and pruned corridors}$ \\
 \textit{// Formulate collision constraints (Sec.~\ref{sec:constraint})} \\
    \ForEach{$s \in \mathcal{S}$}{
        $\mathcal{Z}_{\text{collision}} \gets \text{ConstraintFormulation}(\mathcal{D}^s_{0:N})$
    }

\textit{// Generate contingency plans (Sec.~\ref{sec:BMPCC})} \\
$\vec{z}_{0:N} \gets \text{BMPCC}(\vec{z}_{0}, \mathcal{Z}_{\text{collision}}, \{\pi_s | s \in \mathcal{S} \},  k_b, \mathcal{P}_{ref})$ \\

\Return{$\vec{z}_{0:N}$}
\end{algorithm}
\setlength{\textfloatsep}{8pt}
\section{Maximum Decision Postponing Consideration}\label{sec:adp}

The Branch MPC assumes that by the branching time $k_b$, the uncertainty regarding which predicted mode will occur resolves. At this point, the AV commits to one branch in the scenario tree. Too short branching times risk committing to the wrong branch due to the remaining uncertainty. Therefore, the branching time must be adapted based on the observed situation and the expected resolution of uncertainty. We use the method in \cite{bouzidi2024motionplanninguncertaintyintegrating} which estimates the branching time adaptively by evaluating the overlap of predicted modes to determine when to commit  with high confidence.
However, waiting too long for this uncertainty to resolve can result in missed opportunities (e.g., merging into a gap) possibly leaving no feasible solution. 
 In our case, the selected driving corridors may no longer be feasible after branching. 
Even if there is an overlap between the driving corridor this only means that at this timestep there is one or more positions where the AV can be in both driving corridors at this time. However, this doesn’t imply that it is feasible for the AV to remain within both corridors in future timesteps, as maintaining one corridor may require a different velocity profile than the other (e.g. accelerating vs. braking),  potentially exceeding acceleration or deceleration limits.
The last timestep in which these states can remain equal i.e. delay branching while still ensuring to reach both driving corridors in future is the maximum feasible branching time $k^{max}_{b}$.

To calculate this  time, let us first consider only the longitudinal bounds of two driving corridors.$\mathcal{D}^{>}_{0:N}$ denotes the longer corridor with the higher final progress and $\mathcal{D}^{<}_{0:N}$ denotes the shorter. If we have more than two branches, the driving corridors are first ordered based on final progress. The calculation is then done between the first and last driving corridor.
Let us define the minimum velocity required at timestep \( k \) for the AV to reach at least the lower bound of the longer driving corridor at the next timestep \( \theta^{>}_{k+1, \text{min}} \), given the current progress \( \theta^{>}_{k} \), as:
\begin{equation}
v^{>}_{k, \text{min}}(\theta^{>}_{k}) = \frac{\theta^{>}_{k+1, \text{min}} - \theta^{>}_{k} - 0.5 \cdot a_{\text{max}} \cdot \Delta t^2}{\Delta t}
\end{equation}
Similarly, the maximum velocity $v^{<}_{k, \text{max}}(\theta^{<}_{k})$  to remain within the upper bound of the shorter corridor at the next timestep \( \theta^{<}_{k+1, \text{max}} \), given the current progress \( \theta^{<}_{k} \) can be calculated by applying the minimum acceleration $a_{min}$ instead.

Ensuring these bounds can be reached $\forall k$ is the minimum condition to stay feasible.
By backpropagating these we determine the required position velocity combination at preceding timesteps that satisfies all constraints (for both corridors, reachable progress and velocity at respective timestep and backpropagated constraints). This is done with the inverted model $[\theta_k, v^p_k]^\top = \vec{A}^{-1} ([\theta_{k+1}, v^p_{k+1}]^\top  -  \vec{B} a_{\text{max/min}}) $.

For the longer driving corridor, maximum acceleration is applied, while for the shorter corridor, minimum acceleration is used.  The first timestep where there is at least one state meeting these conditions is identified as the maximum possible branching time $k^{max}_{b,longit}$ (for longitudinal only).

After consideration of the longitudinal part, the lateral motion is considered by evaluating the overlap of driving corridors for  $k^{max}_{b,longit}$ and the subsequent $ k_{\text{lc}} $ timesteps. If no overlap exists, the maximum branching time is reduced until lateral feasibility can be achieved.

The identified maximum possible decision postponing time $k^{max}_{b}$ is then compared to the adaptively estimated decision postponing time. If it exceeds the maximum, one of the selected corridors with the greatest cumulative area is replaced with a backup corridor covering the same mode. Backup corridors are typically more conservative and accommodate higher uncertainty (e.g. mode 1 in fig.\ref{fig:P2}).  If the newly selected corridor still results in to a low maximum branching time $k^{max}_{b}$ or no alternative corridor exists, the branching time $k_b$ is limited to the calculated maximum. The procedure is summarized in Algorithm 2.

\begin{algorithm}[ht]
\caption{Max. Branching Time Consideration}
\SetAlgoLined
\SetKwInOut{Input}{Input}
\SetKwInOut{Output}{Output}

\Input{
  Clustered corridors $\mathcal{D}^{c,A}_{0:N}$ $\forall$ cluster c, \\
  Backup corridors $\mathcal{D}^{m,B}_{0:N}$ $\forall$ modes $m$
}
\Output{
  Updated corridors $\mathcal{D}^{c,A}_{0:N}$, \\
  Branching time $k_b$
}

$k_b \gets \text{GetAdaptiveDecisionPostponing}(\mathcal{D}^{c,A}_{0:N} \ \forall c)$ \\
$k_b^{\max} \gets \text{GetMaximumDecisionPostponing}(\mathcal{D}^{c,A}_{0:N} \ \forall c)$ \\

\If{$k_b > k_b^{\max}$}{
    $c_{\text{remove}} \gets \arg\max\limits_{c} \text{Area}(\mathcal{D}^{c,A}_{0:N})$ \\

    \ForEach{$m \in c_{\text{remove}}$}{
        \If{$\mathcal{D}^{m,B}_{0:N} \text{ is empty}$}{
            $k_b \gets k_b^{\max}$; Break \textit{// as no backup exists} \\
        }
        \Else{
            Remove $c_{\text{remove}}$ from $\mathcal{D}^{c,A}_{0:N}$ \\
            Append $\mathcal{D}^{m,B}_{0:N}$ to $\mathcal{D}^{c,A}_{0:N}$
        }
    }
}
\Return{$\mathcal{D}^{c,A}_{0:N}, k_b$}

\end{algorithm}
\setlength{\textfloatsep}{8pt}
\section{Performance Evaluation} \label{sec:Results}
In this section, we demonstrate the effectiveness of our approach using an intersection scenario and further evaluate it quantitatively on Monte Carlo simulations of random merging scenarios. For simulating the TP behavior we use the Intelligent Driver Model (IDM). 
We compare our method with following baselines:
Conventional MPCC \textbf{(CMPC)} \cite{brito1} plans a trajectory based on a single (the most likely) TP prediction.
Branch MPCC without Scenario Selection \textbf{(BMPC noSS)} considers the four most likely TP predictions without scenario selection.
Branch MPCC variant in\cite{bouzidi2024motionplanninguncertaintyintegrating} \textbf{(BMPC UVD)}  selects the two most relevant  TP predictions based on topology criteria using uniform deformation visibility(UVD).
Reach Scenario-based MPCC \textbf{(RSMPC)} combines our scenario selection with conventional Scenario-based MPC \cite{kensbock_scenario-based_2023}. This Baseline constructs a single driving corridor using the 4 most likely predictions.
Reach Branch MPC without Maximum Decision Postponing \textbf{(RBMPC noMaxDP)} is an ablation of our method not using the maximum decision postponing calculation. 
Further, we evaluate our method \textbf{RBMPC} with both two and three selected scenarios/corridors. If  the number of  relevant scenarios output by our approach in Sec.\ \ref{sec:scen} is higher than the number of considered scenarios, we select the ones with the highest cumulated probability of the respective predictions.
All Branch MPC-based baselines employ the Motion Transformer\cite{shi_motion_2022} which outputs 6 different predictions for each TP and  the adaptive decision postponing method introduced in \cite{bouzidi2024motionplanninguncertaintyintegrating}. We implement all approaches using CasADi with the IPOPT solver, utilizing a prediction horizon of \(N = 40\) and a sampling time of \(\Delta t = 100\text{ ms}\).

\begin{table*}[t]
    \vspace{0.5em} 
    \caption{Monte Carlo Analysis of 100  random gap merging scenarios with 3 TPs}
    \vspace{-0.6em} 
    \centering
    \begin{tabular}{ |p{4.15cm}||m{1.2cm}|m{1.2cm}|m{1.25cm}|m{1.2cm}|m{1.2cm}|m{1.2cm}|m{1.2cm}|m{1.15cm}|  }
 \hline
  & \multicolumn{3}{c|}{\textbf{Merging Execution}} & \multicolumn{5}{c|}{\textbf{Mean Performance}} \\ 
 \cline{2-9}
  \rule{0pt}{8pt} & Success $\uparrow$ & Aborted $\downarrow$ & Collision $\downarrow$&  $\bar{v}[m/s] \uparrow $ & $\bar{|j|}[m/s^3]\downarrow$& $|\bar{\delta|}[rad]\downarrow$& $\bar{d}_{min}[m]\uparrow$ &$\bar{t}_{c} $ $[ms]\downarrow$ \\ 
 \hline
 \textbf{CMPC (1 Scen.)} & 87\% & 7\% & 6\% &  6.84  & 1.63 & \textbf{0.05} &4.31 & \textbf{36} \\ 
  \hline
   \textbf{BMPC noSS (4 Scen.)} & 92\% & 7\% & 1\% & 7.76   & 1.35 & \textbf{0.05} &\textbf{7.46} & 233\\
  \hline
\textbf{BMPC UVD (2 Scen.) } & 92 \% & 6 \% & 2 \% & 7.57  & 1.42 & 0.06 &5.77 & 88+17\\ 
  \hline
 \textbf{ RSCMPC (4 Scen.)} & 87 \% & 13 \% & \textbf{0\%} & 6.40  & 1.28 & 0.08 & 7.33&  23+41 \\ 
  \hline
   \textbf{BMPC noMaxDP (2 Scen.) (Ours.)} & 94 \% & 6 \% & \textbf{0\%}&  \textbf{8.18} & 1.08 & 0.06 &6.40 &31+33 \\ 
  \hline
   \textbf{RBMPC (2 Scen.) (Ours.)} & \textbf{96 \%} &  \textbf{4\%} &  \textbf{0\%} &  8.17 & \textbf{1.06} & 0.06 & 6.81 & 31+33 \\ 
  \hline
 \textbf{RBMPC (3 Scen.) (Ours)} & \textbf{96 \%} & \textbf{4\%}   & \textbf{0\%}&  \textbf{8.18}  & 1.13 & 0.07 & 7.34 & 61+38\\   
 \hline
\end{tabular}
\label{tab:mca}
\vspace{-1.0em} 
\end{table*}

\subsection{Qualitative Evaluation} 

\begin{figure}[b]
\centering
    \includegraphics[trim = 0mm 0mm 0mm 0mm, clip, width=0.49\textwidth]{"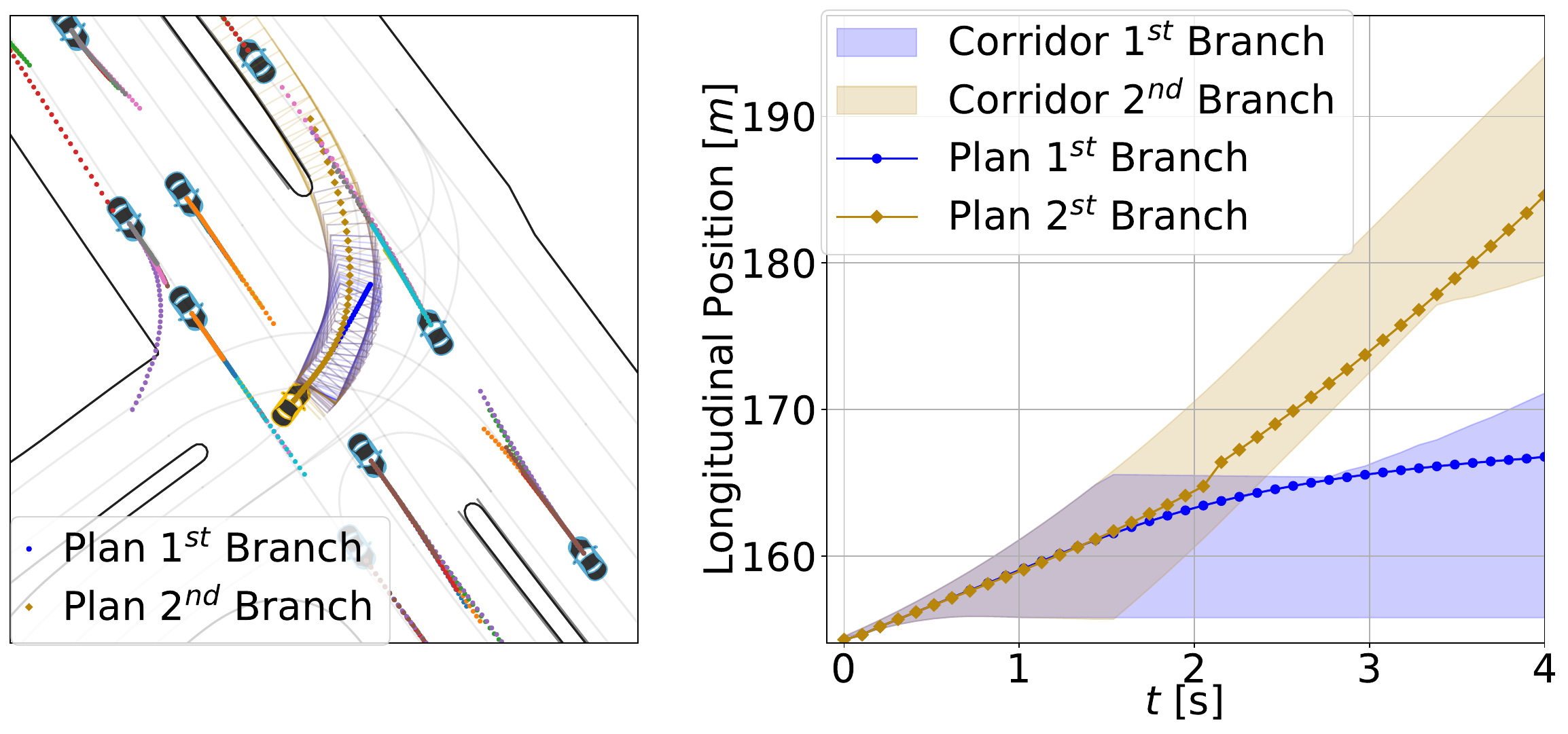"}
    \vspace{-1.3em} 
    \caption{Snapshot with prediction of TPs and the selected driving corridors and respective plans of the AV.}
    \label{fig:P4} 
    \vspace{-0.4em} 
\end{figure}

\begin{figure}[]
\centering
    \includegraphics[trim = 1mm 2mm 0mm 2.5mm, clip, width=0.5\textwidth]{"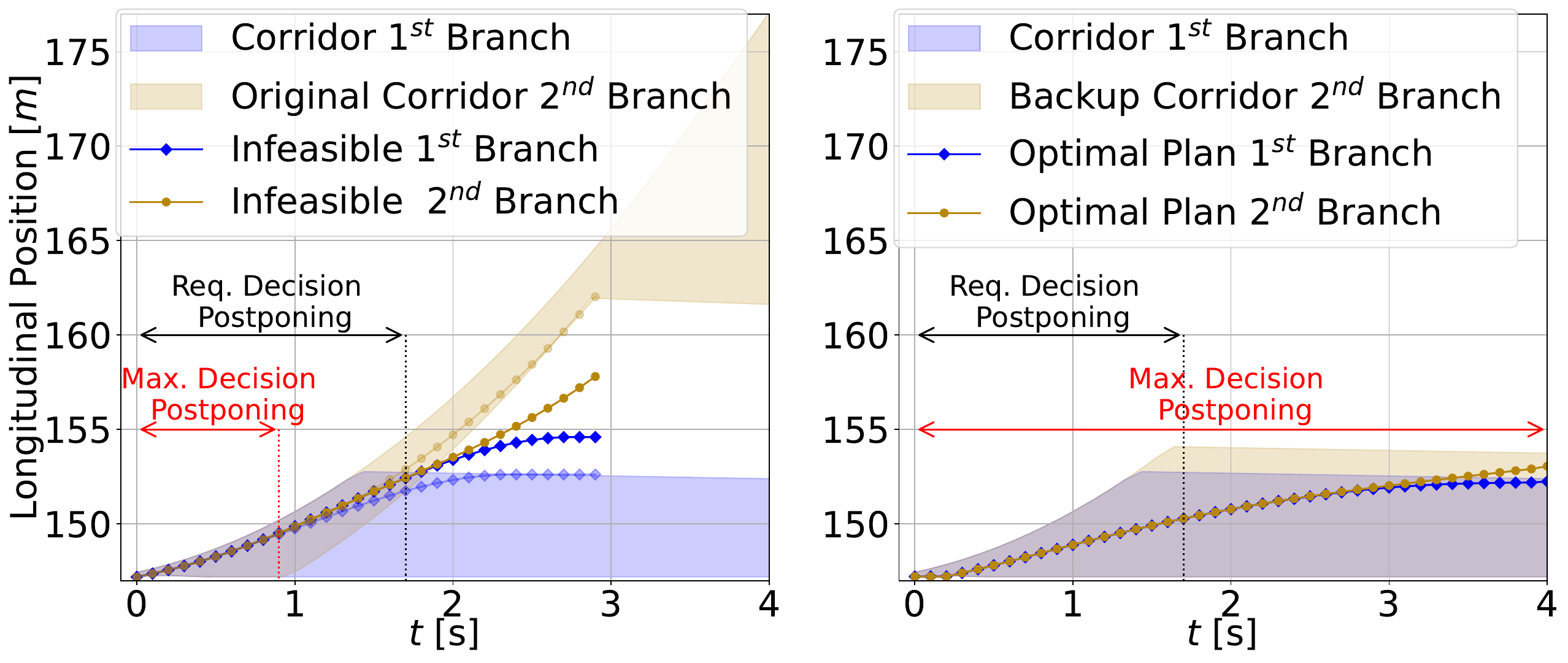"}
    \vspace{-1.3em} 
   \caption{Selected driving corridors and respective maximum feasible decision postponing times before (\textbf{left}) and after (\textbf{right}) replacement to ensure compliance with the required decision postponing time for achieving sufficient certainty.} 
    \label{fig:P5} 
    \vspace{-0.4em} 
\end{figure}

\begin{figure}[htbp]
\centering
    \includegraphics[trim = 0mm 0mm 0mm 3.5mm, clip, width=0.49\textwidth]{"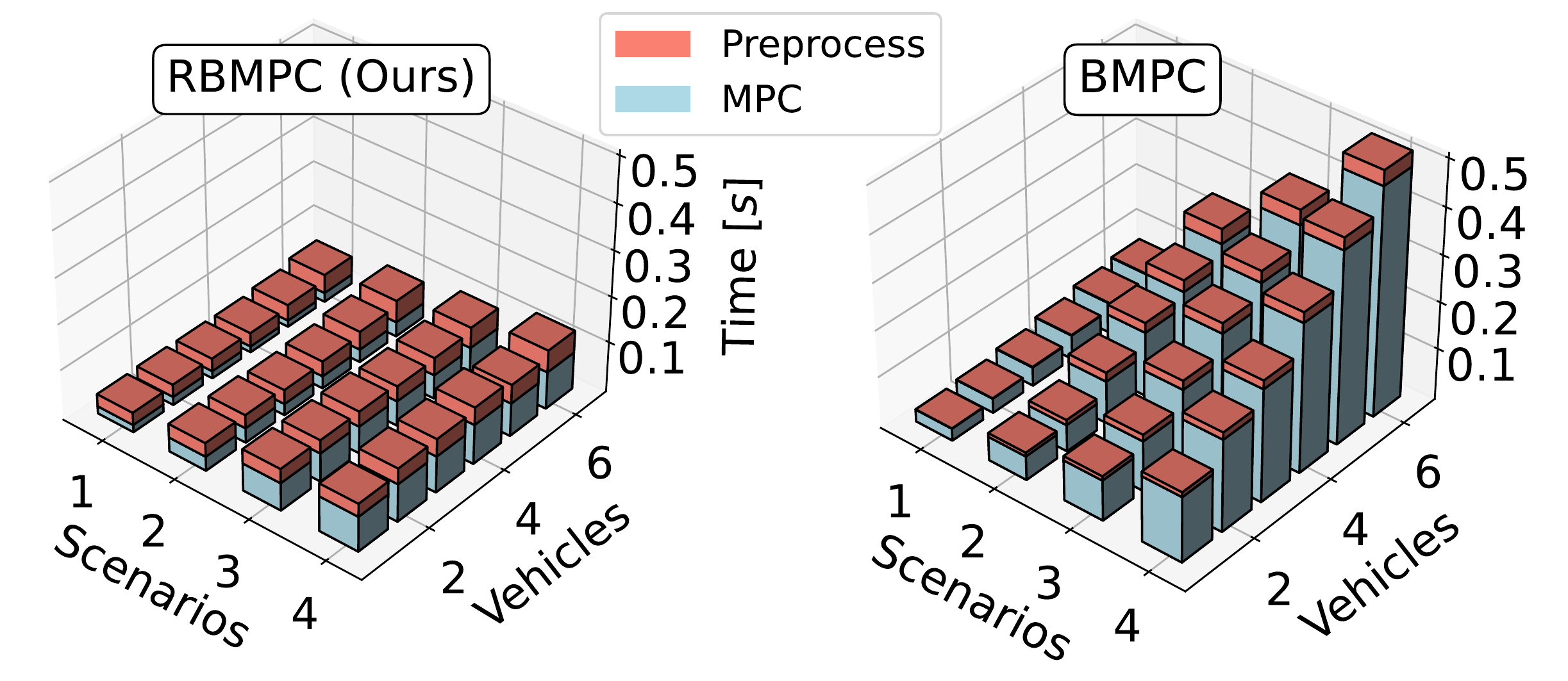"}
    \vspace{-1.3em} 
    \caption{Computation time of our method and BMPC UVD with increasing number of branches and TPs splitted in preprocessing (scenario selection) and MPC solving time.}
    \label{fig:P6} 
    \vspace{-0.4em} 
\end{figure}

We evaluate our method (RBMPC 2 Scen.) in a challenging intersection scene with 12 traffic participants (TP), achieving runtimes under 100 ms (Fig. \ref{fig:P4}). Unlike the Baselines, our method successfully completes the turn by identifying safe driving corridors for the multimodal behavior predictions of the TPs. 
By contrast, CMPC results in a collision due to its neglect of critical predictions, while RSCMPC becomes overly conservative and gets stuck at the entrance of the intersection. The RSCMPC  behavior underscores the significance of branching, since the RSMPC misses potential opportunities to cross as it does not consider future information gain that the AV can react on. Our ablation  method RBMPC noMAXDP also fails to cross the road safely in this scene which highlights why the maximum decision postponing consideration presented in sec. \ref{sec:adp} is essential. The reason for that can be observed in fig. (\ref{fig:P5}) which illustrates a snapshot of the initially selected driving corridors at a certain timestep. Our Adaptive Decision Postponing approach outputs that the required branching time for sufficient expected certainty is 1.7 seconds.  However, the maximum feasible branching time that ensures the AV remains within the branching constraints over the entire prediction horizon is 0.9 seconds. Consequently, our method replaces the higher-progress driving corridor with a more conservative one to accommodate the longer branching time.  Without this adjustment, applying the required branching time to the original corridors leads to infeasible solutions, impairing safety. The remaining Branch MPC Baselines are not capable to run in real-time with high number of TPs.  
To demonstrate the computational scalability of our approach compared to other Branch MPC methods, we evaluate it alongside BMPC UVD across multiple intersection scenes with varying numbers of TPs and  branches. Fig. (\ref{fig:P6}) shows that our method significantly outperforms BMPC UVD in computation time, especially as the complexity increases.

\subsection{Quantitative Evaluation} 
We quantitatively compare our method to the baselines in Table (\ref{tab:mca}) using 100 Monte Carlo simulations with randomly generated merging scenarios. For these simulations, we randomly sample IDM parameters and the initial states of the vehicles. Beyond evaluating the success rate of the merging maneuvers, we assess metrics such as average velocity (\( \bar{v} \)), average absolute jerk (\( \bar{|j|} \)), average absolute steering (\( \bar{|\delta|} \)), computation time (\( \bar{t}_c \))—split into preprocessing time (e.g., scenario selection) and MPC solving time (\( t_{\text{MPC}} \)), and the average minimum distance (\( \bar{d}_{min} \)) across all scenarios.
\begin{figure}[!htbp]
\centering
    \includegraphics[trim = 46mm 152mm 6mm 190mm, clip, width=0.55\textwidth]{"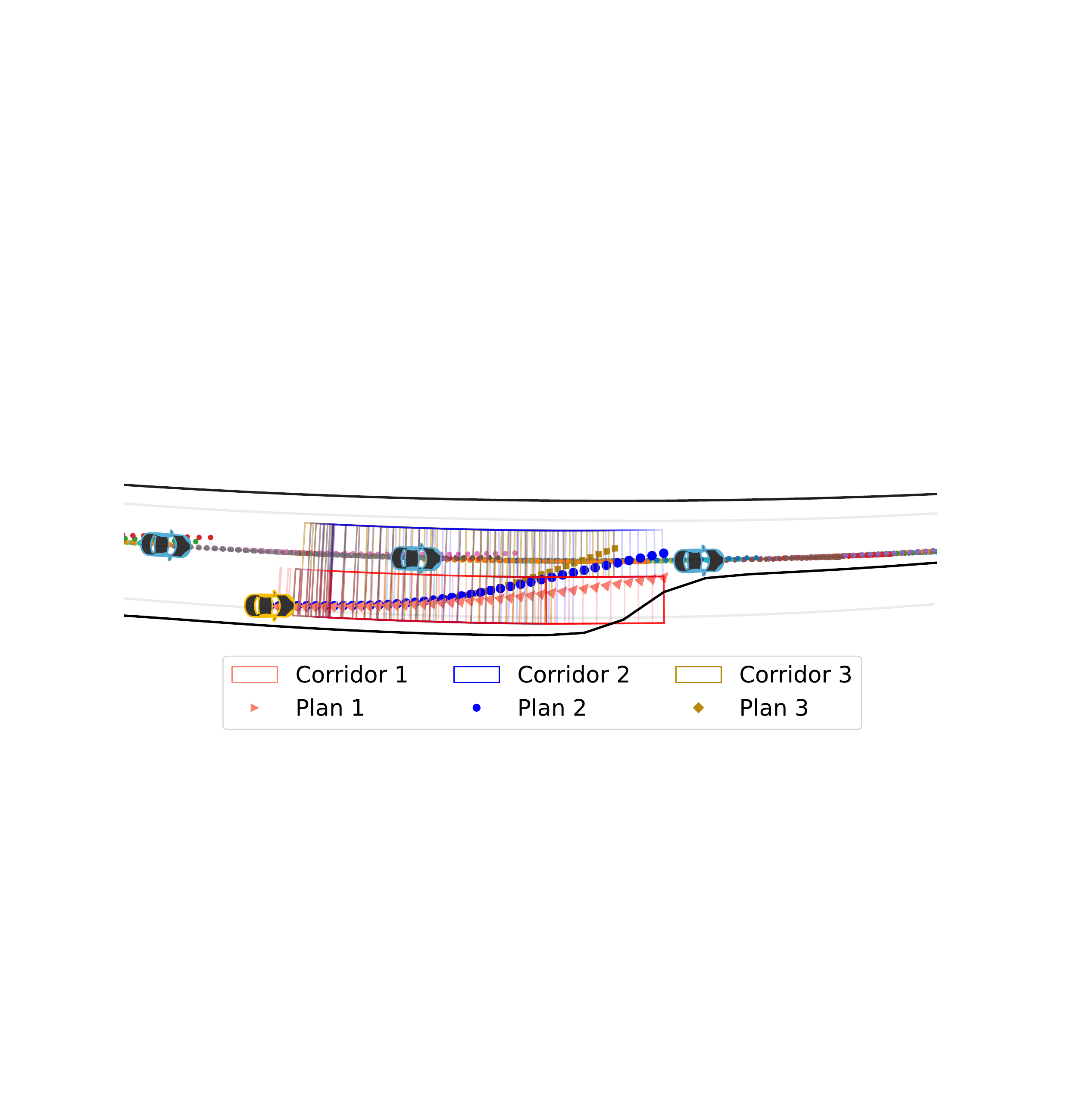"}
    \vspace{-1.0em}
    \caption{ Snapshot of our method during merging outputting three branches/driving corridors with their respective plans}
    \label{fig:P7} 
    \vspace{-0.6em} 
\end{figure}
The results in Table (\ref{tab:mca}) demonstrate that safety is significantly improved when multiple predictions of the traffic participants (TP) are considered. CMPC, which neglects such predictions, consistently exhibits the worst performance. The table also highlights the importance of incorporating future feedback: RSCMPC, being overly conservative, results in a high number of aborted maneuvers.  

\begin{figure}[!htbp]
\centering
    \includegraphics[trim = 0mm 0mm 0mm 0mm, clip, width=0.49\textwidth]{"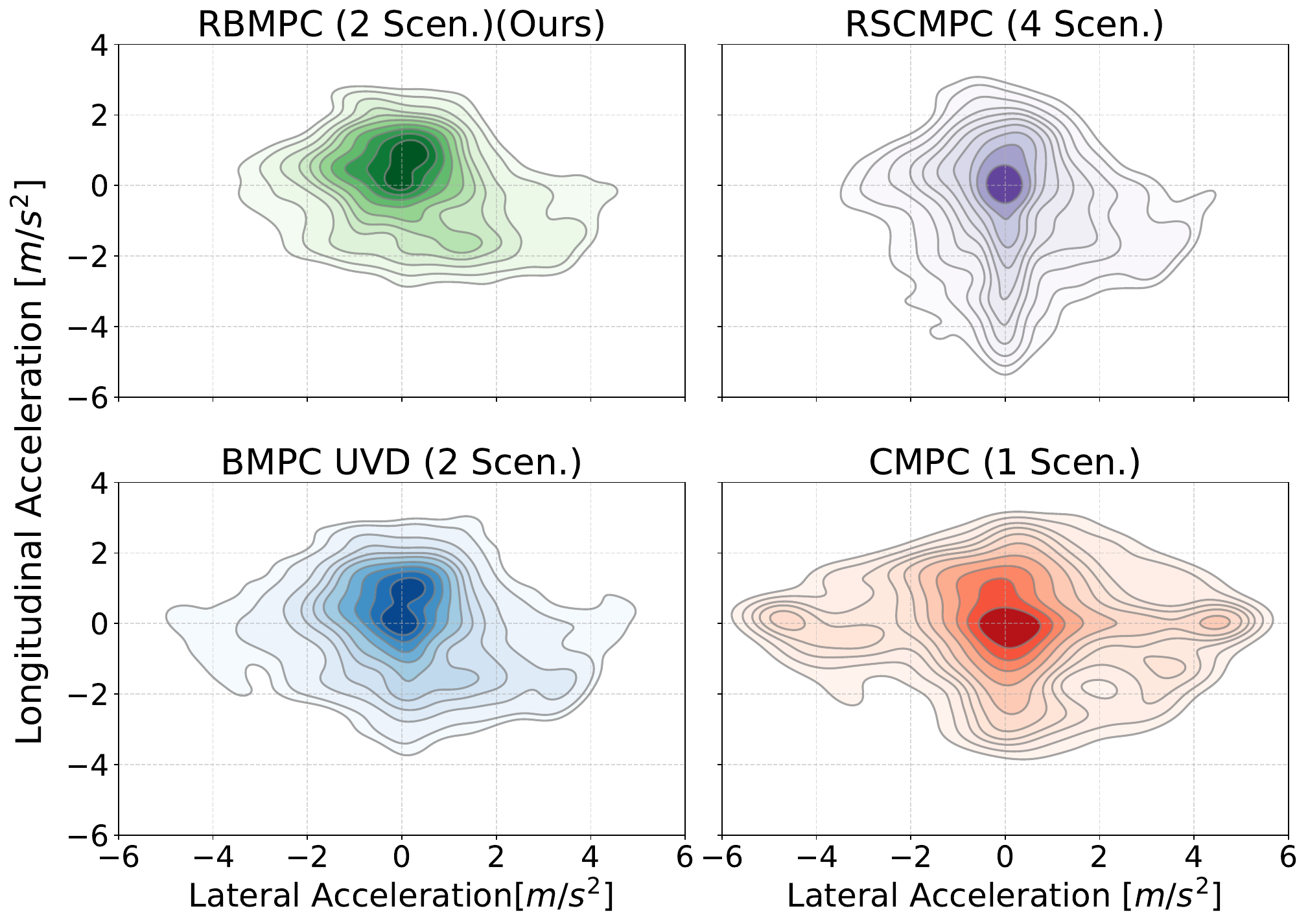"}
    \vspace{-1.3em} 
    \caption{ Distribution of comfort metric lateral and longitudinal acceleration for methods compared in Table \ref{tab:mca}.
}
    \label{fig:P8} 
    \vspace{-0.8em} 
\end{figure}

Our evaluation further reveals that increasing the number of branches in the scenario tree beyond 2 scenarios provides negligible performance gains when using our scenario selection strategy. However, the choice of scenario selection method is critical, as evidenced by the performance gap between BMPC UVD and RBMPC. With our reachability-based scenario selection, we achieve higher safety and comfort while maintaining lower computation times. 
To complement these metrics, we plot kernel density estimates of longitudinal and lateral accelerations of all merging scenes in Fig.\ \ref{fig:P8}. While averages alone may not fully capture performance nuances, these plots show that our method results in fewer cases outside the comfortable human driving zone recorded in\cite{unknown}.
\vspace{-0.1em}


\section{Conclusion}\label{sec:conclusion}
We introduced a planning framework that integrates learning-based multi-modal predictions with Branch MPC through reachability analysis. The scenario tree of the Branch MPC is build based on how each predicted behavior influences the AV using driving corridors. This  addresses the scalability and safety challenges arising from the scenario tree construction with multi-modal traffic uncertainty. The framework demonstrates excellent scalability with increasing traffic participants, making it suitable for dense traffic. While increasing the branches in the scenario tree is possible, it does not substantially improve performance, as most predictions are effectively handled by a minimal number of driving corridors. Through our pruning process, each driving corridor ensures constraint satisfaction for all included modes.  The integration of maximum feasible decision postponing balances the benefits of passive information gathering against the risk of missing critical maneuver opportunities. Experimental results confirm that our framework significantly improves both safety and comfort while maintaining real-time performance.

\bibliographystyle{IEEEtran}
\bibliography{Refs}
 
\end{document}